\documentclass[useAMS,usenatbib]{mn2e}

\usepackage{epsfig}
\usepackage{color}
\usepackage{soul}
\usepackage{amssymb}
\title[Practical suggestion on detecting exomoons]
{Signals of exomoons in averaged light curves of exoplanets}
 
\author[A. E. Simon,  Gy. M. Szab\'o, L. L. Kiss and K. Szatm\'ary]{
A. E. Simon$^{1,2}$\thanks{E-mail: atthys@konkoly.hu},
Gy. M. Szab\'o$^{1,2}$\thanks{E-mail: szgy@konkoly.hu},
L. L. Kiss$^{1,3}$
and K. Szatm\'ary$^{2}$ \\
$^1$Konkoly Observatory of the Hungarian Academy of Sciences, PO. Box 67, H-1525 Budapest, Hungary\\
$^2$Department of Experimental Physics and Astronomical Observatory, University of Szeged, 6720 Szeged, Hungary\\
$^3$Sydney Institute for Astronomy, School of Physics A28, University of Sydney, NSW 2006, Australia
}

\begin{document}

\date{Accepted Received; in original form}

\pagerange{\pageref{firstpage}--\pageref{lastpage}} \pubyear{2010}

\maketitle

\label{firstpage}

\begin{abstract}

The increasing number of transiting exoplanets sparked a significant interest in discovering their moons. 
Most of the methods in the literature utilize timing analysis of the raw light curves. 
Here we propose a new approach for the direct detection of a moon in the transit light curves 
via the so called Scatter Peak. The essence of the method is the evaluation of the local scatter in the folded
light curves of many transits. We test the ability of this method with different simulations:
Kepler ``short cadence'', Kepler ``long cadence'', ground-based millimagnitude photometry with 3-min cadence,
and the expected data quality of the planned ESA mission of PLATO.
{The method requires $\approx$100 transit observations, therefore applicable for moons of 10-20 day period planets,
assuming 3-4-5 year long observing campaigns with space observatories.}
The success rate for finding a 1 $R_{Earth}$ moon 
around a 1 $R_{Jupiter}$ exoplanet turned out to be quite promising even for the simulated 
ground-based observations, while the detection limit of the expected PLATO data is around 0.4 $R_{Earth}$.
We give practical suggestions for observations and data reduction to improve the chance of such a detection:
(i) transit observations must include out-of-transit phases before and after a transit,
spanning at least the same duration as the transit itself;
(ii) any trend filtering must be done in such a way that the preceding and 
following out-of-transit phases remain unaffected. 

\end{abstract}


\section{Introduction}

The number of known transiting exoplanets is rapidly increasing, which
has recently inspired significant interest as to whether they can host a
detectable moon. Although there has been no such example where the
presence of a satellite was proven, several methods have already been
investigated for such a detection in the future. The most important
methods evaluate the timing of transits, e.g. barycentric Transit
Timing Variation, TTV Sartoretti \& Schneider 1999, Kipping 2008,
photocentric Transit Timing Variation, TTVp, Szab\'o et al. 2006, Simon
et al 2007, Transit Duration Variation, TDV, Kipping, 2009,
Time-of-Arrival analysis of pulsars, Lewis et al. 2008). {There are
further photometric methods for observing rings of exoplanets (Ohta et al. 2009, Di Stefano et al. 2010) 
starspots in transits (Silva 2003, Silva et al. 2010 and references therein), or even, transits of alien spacecrafts (Arnold, 2005).}

Here we propose a photometric method for the detection of the moon directly in the
raw transit light curves. When the moon is in transit, it puts its own
fingerprint on the intensity variation. In realistic cases, this distortion is
too little to be detected in the individual light curves. Simply taking the
boxcar average of a folded light curve that consists of many transits, is not a
powerful solution because it results in a significant amount of correlated
(``pink'') noise. The smooth variation of this correlated noise can mimic/hide
the real distortions of the light curve due to the moon. 
Here we introduce the scatter of the folded light curve as an appropriate estimator
for the presence of a moon. The stability of the method relies on its robust
nature, i.e. the scatter will be estimated in a boxcar that is comparable,
or even longer, than the transit duration. 

 {Here we show that a careful analysis
of the scatter curve of the folded light curves enhances the
chance of detecting the exomoons directly. Our aim is to present
a detection technique that is very specific, i.e. when the
test is positive, the presence of an exomoon is probable.
With careful pre-processing of the light curves (e.g., by recentering 
the transits) signals that can mimic exomoons are largely suppressed. 
Consequently, the Scatter Peak method can be considered both as a tool 
(i) for quickly finding systems that warrant more detailed analyes and 
(ii) for confirming the presence of an exomoon when suspected from TTV and/or TDV analyses.}

\section{A simple model for the averaged light curves}

\subsection{Averaged transit light curves with a moon}

\begin{figure}
\centering
\includegraphics[width=8cm]{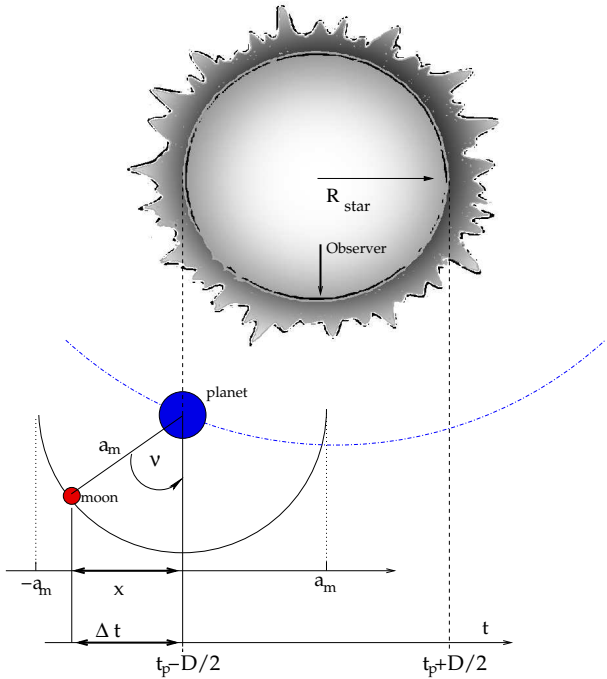}
\caption{Transit geometry of a star--planet--satellite system.}
\end{figure}

 {Here we describe a very simple model to illustrate the concept of the Scatter Peak. 
For the sake of clarity, we consider a special configuration (Fig. 1)
that can be handled analytically. Generic transit light curves with a moon
will be examined later in numerical simulations.}

Let us assume a moon on circular orbit, and therefore, we {\it a
priori} know the shape of the light curve component of the moon (it is similar to that
of the planet in shape and duration, but with shallower transit). In this case,
the orbital inclination of the moon is 90$^\circ$. We assume further that the moon orbits slowly, i.e. the moon -- planet geometry does not change significantly during the transit while 
the transits of the moon appear somewhat earlier or later than the planet's transit.
We also make use of the knowledge that TTV was initially removed from data by transforming the time
(i.e. shifting the derived transit to the predicted value by recentering, Sect. 4.1).

Let $f(x)$ be the density function of $x$, the moon's projected position. Here $x:=a_m*\sin\nu$, where $\nu$ is the anomaly  and $a_m$ is the semi-major axis of the moon (Fig. 1). Because the orbit is circular, $\nu$ follows uniform distribution. After some calculus (see Appendix for the details) we get the density function of $x$ which is
\begin{equation}
f(x) dx \propto {1 \over \sqrt{ a_m^2 - x^2}} dx.
\end{equation}
{\it The projected distribution of the moon around the planet follows an $1/\sqrt{a_m^2-x^2}$ distribution.} 

Because both the transit light curve of the moon, $lc(t)$ and the averaged light curve is a function of time, $f(x) dx$ must be rewritten to time domain. If the projected position of the moon is $x$ apart from the planet, the time delay
between the transits of the moon and the planet is $x/v_{orb}$, where $v_{orb}$ is the orbital velocity of the planet.
The appropriate transformation to the time domain is therefore $\Delta t=x/v_{orb}$. Here the relative transit time of the moon $\Delta t =t-t_p$, where $t_p$ is a time of the transit of the planet. With this notation, the distribution of the transit time of the moon is
\begin{equation}
f(\Delta t) dt \propto {1\over \sqrt{(a_m/v_{orb})^2 - (\Delta t)^2}} dt.
\end{equation}

In every case, when the moon is observed in an individual light curve,
the occulted flux will be the sum of the transit light curve of the
planet centred at $t_p$, and the transit light curve of the moon, centred at $t_p+\Delta t$. The light curve of a single event is a convolution of the transit light curve with two Dirac delta functions with different weights, representing the planet and the moon at $t_p$ and $t_p+\Delta t$. The average light curve of many events is the expectation flux, taking all $\nu$ values into account. At this step, the planet component can be subtracted, and the residual of the moon will remain alone. Since we average many events, the many individual Dirac delta functions representing the moon will follow the distribution of $f(\Delta t)$, thus the many delta functions in the summation can simply replaced by a convolution with $f(\Delta t)$. Consequently, the $lc_m(\Delta t)$ light curve components due to the moon will be averaged to $\overline{lc_m(\Delta t)}$, which is
\begin{equation}
\overline{lc_m(\Delta t)} = f(\Delta t) \otimes lc(\Delta t),
\end{equation}
where $\otimes$ represents a convolution.


\subsection{The Scatter Peak}

\begin{figure*}
\begin{centering}
\epsfig{file=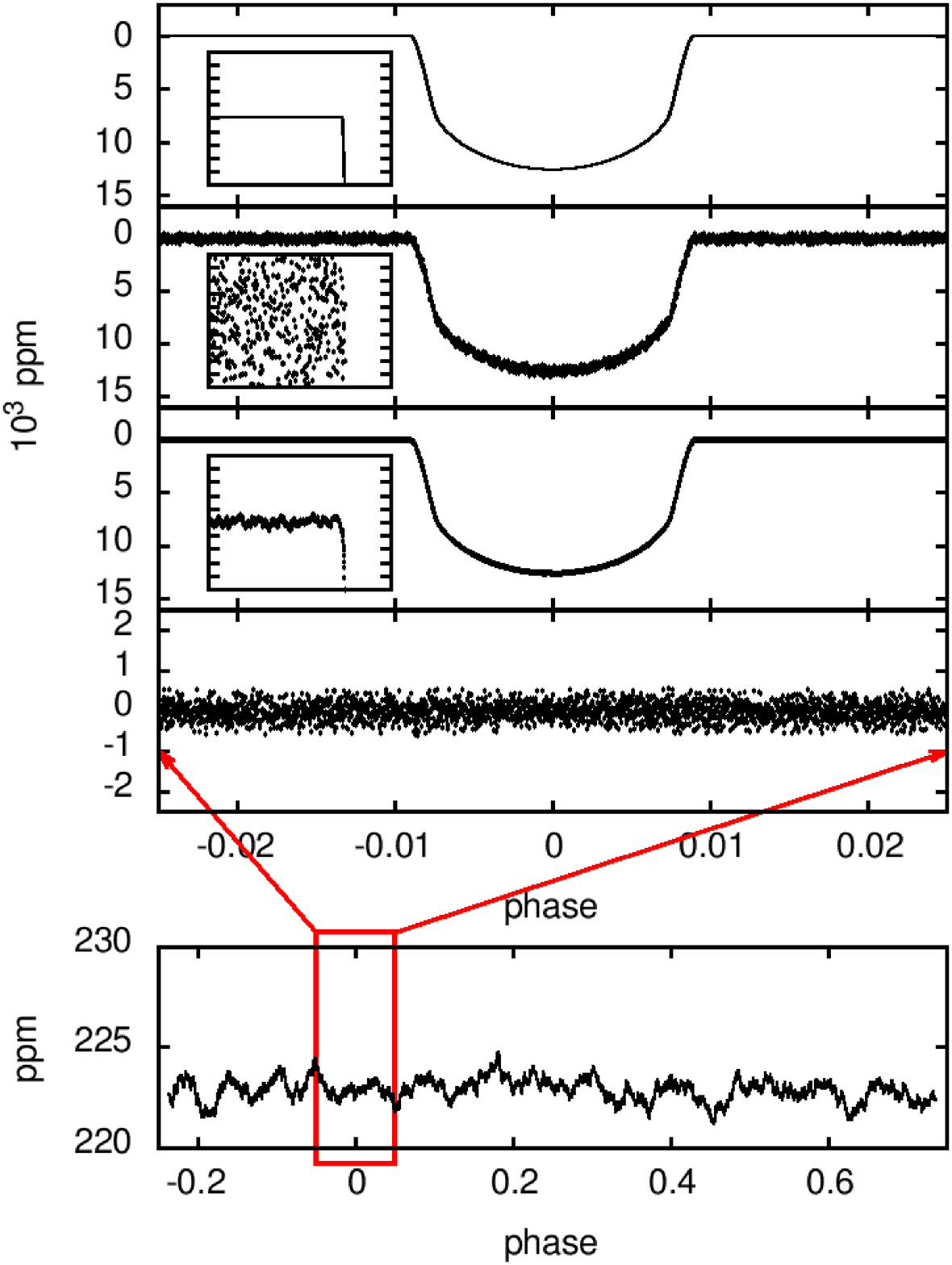,width=8cm}
\hskip 1.5cm 
\epsfig{file=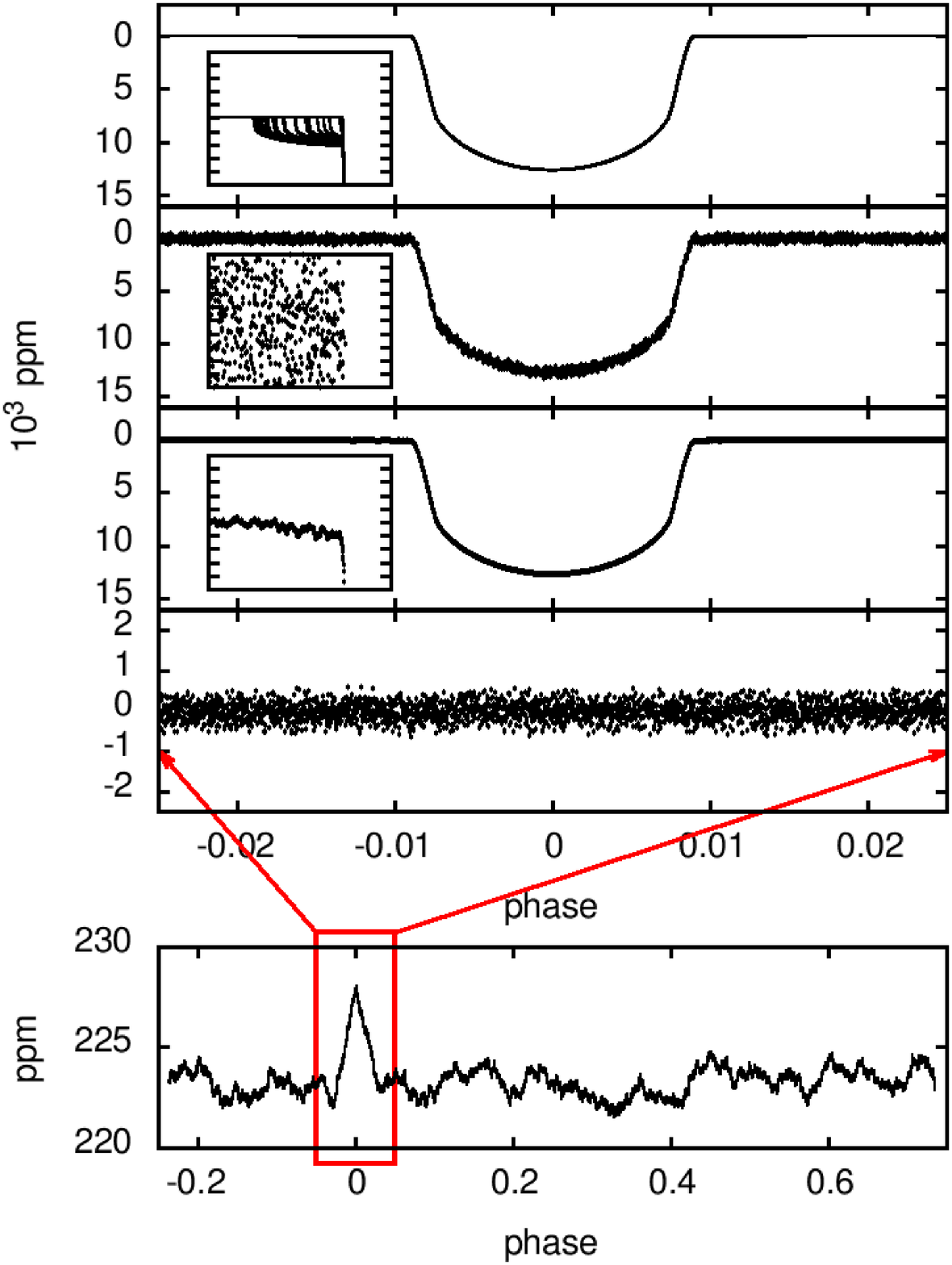,width=8cm}
\caption{{Simulations of 109 transits of an 1 $R_{Jupiter}$ size planet with}
Kepler short cadence sampling and noise. Left panel: simulations without a moon; Right
panel: scatter peak of a 1.0 $R_{Earth}$ sized moon. 
Each column shows the input light
curves, the noisified light curves, the median filtered light curves,
the residuals to the median and the rms scatter of the residuals (the
inserts plot the ingress phase of the exomoon; tick-step is 50 ppm).}
\end{centering}
\end{figure*}

The presence of a moon at a given $\Delta t$ transit time follows
a distribution with a local probability defined in Eq. 2. 
Provided that the moon is in fact at the given position, $lc(\Delta t)$, the light curve component
associated to the moon will be known everywhere. From a set of $\Delta t$ positions distributed according to Eq. 2, one can infer the successive distribution of light loss at each generic time $\tau$. In the general case, this distribution will be of the multinomial family, with
a non-trivial shape (i.e. if the transit parameters are such that ingress-egress phases are shallower/steeper, there will be more/less probability to detect just little light hidden by the moon). Of course simulations can easily support parameter-dependent distributions, but for the theoretical framework it is more prudent to consider a very simple light curve 
shape: a simple box with the duration ($D$) and the depth ($\delta m_{moon}$) as free parameters.

Within this framework, the light occulted by the moon at a generic $\tau$ time will be equally $\delta m_{moon}$ if the moon's position is closer to $\tau$ than $D$ ($|\Delta t-\tau|<D/2$), and will be equally zero elsewhere. Hence we know the distribution of the moon itself, this condition on the relative positions can be evaluated, leading to $\delta m_{moon}$ light is occulted by the moon with a probability $p$ expressed by a convolution
\begin{equation}
p(\Delta t)=\int_{\tau-D/2}^{\tau+D/2} \! \! \! \! f(\Delta t) d\tau \ \ \equiv \ \ f(\Delta t) \otimes I(|\Delta t-\tau|<D/2)
\end{equation}
at a generic time $\tau$. Here $I(C)$ is the identity function that is $1$ whereas $C$ is true and $0$ elsewhere. With this formulation,
the light curve components due to the moon will be binomially distributed, and the local
scatter of light curves can be estimated using the standard deviation of the binomial distribution
\begin{equation}
rms = \delta m_{m} \sqrt{p(1-p)},
\end{equation}
which is the scatter curve of the moon's transit, 
and $\delta m_{m}$ is the expected light loss if the single moon is in transit. 

Because precise measurement of scatter requires the analysis of many data points, 
the light curves will be evaluated in a very wide long boxcar in practice, 
to derive precise scatter values. 
This sampling will behave as a convolution kernel acting on the local values of the scatter,
and finally, the shape of scatter curves will be reduced to a simple
wide peak (the ``Scatter Peak'') around the transit time of the planet. 

In the followings, we examine the Scatter Peak in real simulations.

\section{Simulations} 

 {We made realistic numerical simulations with our image-level simulator 
(Simon et al. 2009) to examine the discovery probabilities of large exomoons
in light curves of different photometric qualities. 
The reliability analysis (Sect. 4.1.) invokes an accurate estimation of 
the light curve scatter and moreover, the scatter of the scatter. Because
these statistical variables are highly fluctuating, there is a demand for $\approx$ 100
transit data for convincing results. Additionally, the same length of data set is required without a moon,
interpreting the null signal event. This will be the reference in making the decision
whether the detection of the moon is significant.
}
 {
Therefore, all subsets incorporated 109 individual transit light curves. 
Continuous datasets are delivered by space observatories and 
even the longest ones from $Kepler$ will span about 3 to 5 years at most. 
Hence, the Scatter Peak is restricted for exomoons around planets with $P_{orb} \lesssim $ 10-20 days.
That is why we selected $P_{orb}$=10 days for the model planet. This is also favourable
because of the distribution of the currently known exoplanets: the majority of them has orbital
periods in this order of magnitude. However, we recall again that the most
relevant parameter is not the period of the planet, but the number of transits which we are
able to observe.
}

 {The planet was a hot Jupiter with 0.7 $M_{Jupiter}$, 1.0 $R_{Jupiter}$ mass and radius on a circular orbit
with $a_{planet}=0.09$~AU.
The model moon orbited at $a_{moon}=6.84\times10^5$~km, 82\%{} of the Hill-radius, and had 
a period of $P_{moon} = 4.3$~days. 
This configuration was considered to generally represent a non-resonant case while the moon was
enabled to orbit during the transit and mutual transits were also comprehended.  
The central star was a solar analogue. Sample light curves of such a system are shown in the top rows of Fig. 2. 
}

Transit light curves of four different qualities were simulated.  
One dataset represented the best 
quality ground-based (GB) photometry with 178 second sampling and 0.23
mmag standard deviation of the light curve points (0.7 mmag error, closely mimicking what has
been achieved by Southworth et al., 2010). 
Space measurements were represented by Kepler space telescope short cadence (SC)
and long cadence (LC) samplings and the bootstrap noise of non-variable
stars.  The quality of future space observatories was represented by
the anticipated data quality of ESA's planned PLATO mission. For the ``PLATO''
quality dataset we assumed 25.13 sec sampling and an accuracy of 0.12 mmag
(data taken from Catala et al. 2011). In such way, 16 subsets of light curves were calculated, each 
representing individual systems with different moons (0.5 to 1.0
$R_{Earth}$ for ground-based, Kepler LC and SC quality, and 0.4 to 1.0 $R_{Earth}$ for 
PLATO quality). 


\section{Detection strategy}
 {The secure detection of a moon relies on four important steps. After pre-processing the data,
the detection parameters have to be fine-tuned, then applied to the observations and finally, we make a decision
on whether the Scatter Peak is significant. The detailed recipe of the entire process is as follows.}

\begin{figure*}
\begin{centering}
~ \hskip0cm{} Kepler LC \hskip4.5cm{} Kepler SC  \hskip4.5cm{}PLATO\\
\epsfig{file=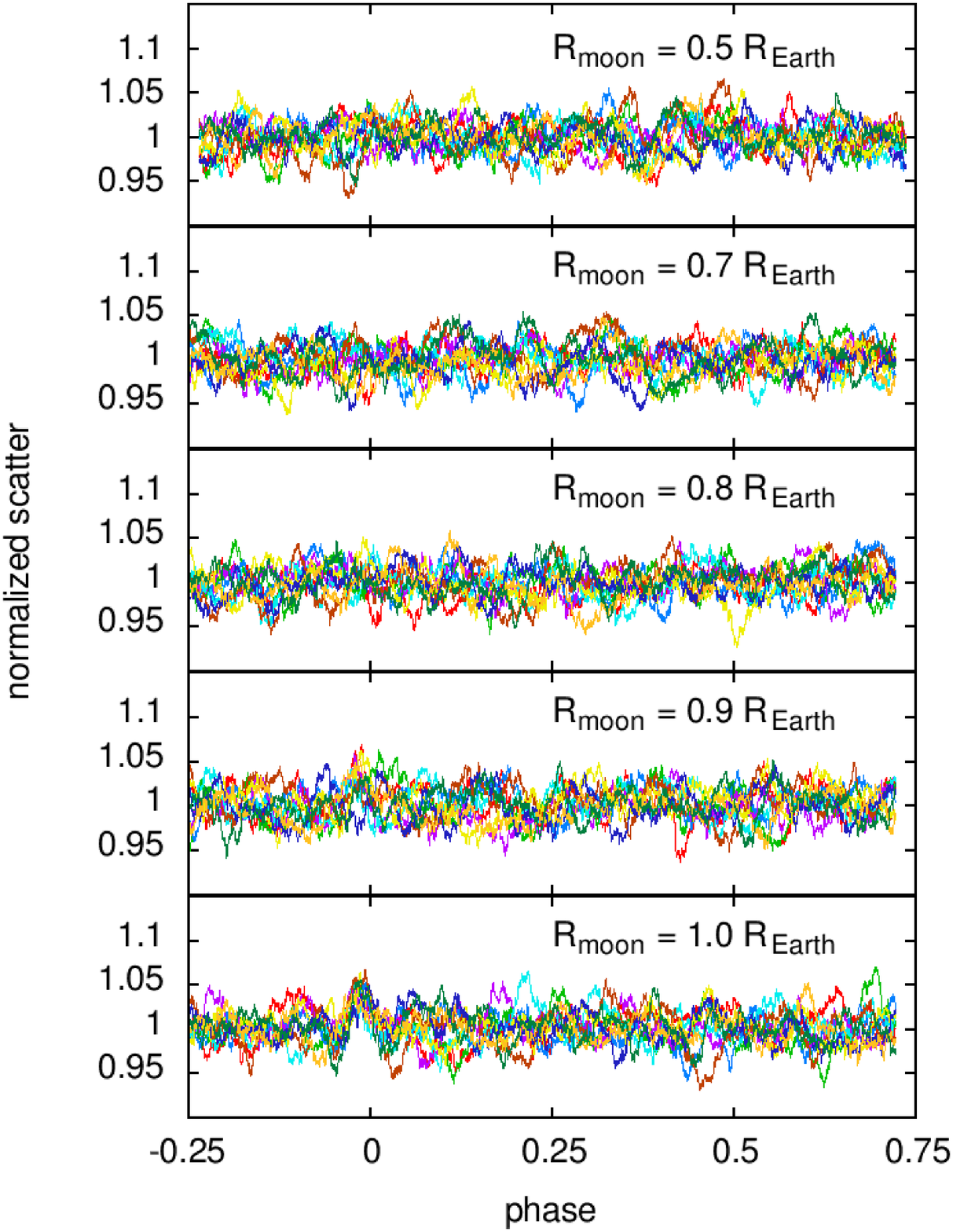,width=5.8cm}
\epsfig{file=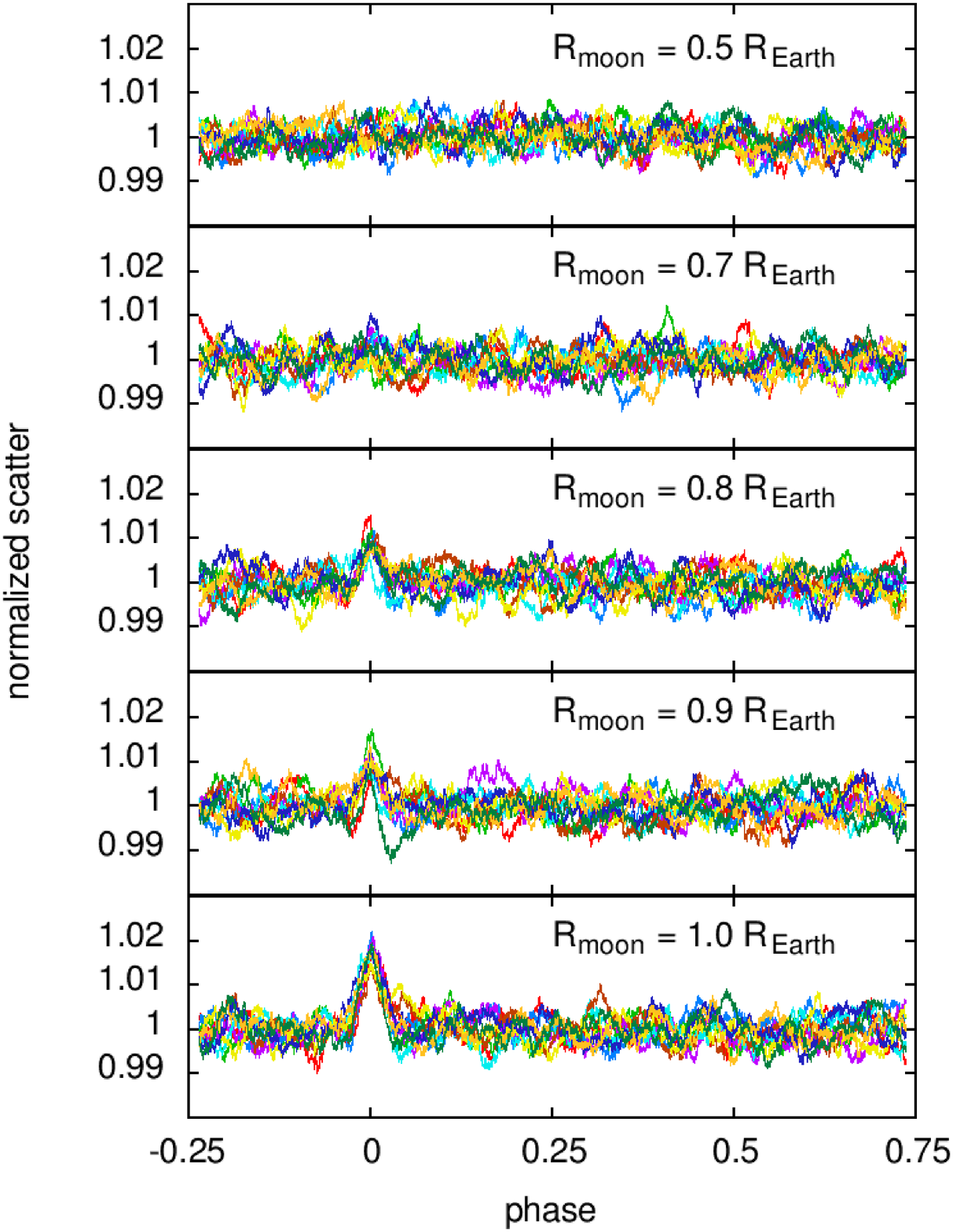,width=5.8cm}
\epsfig{file=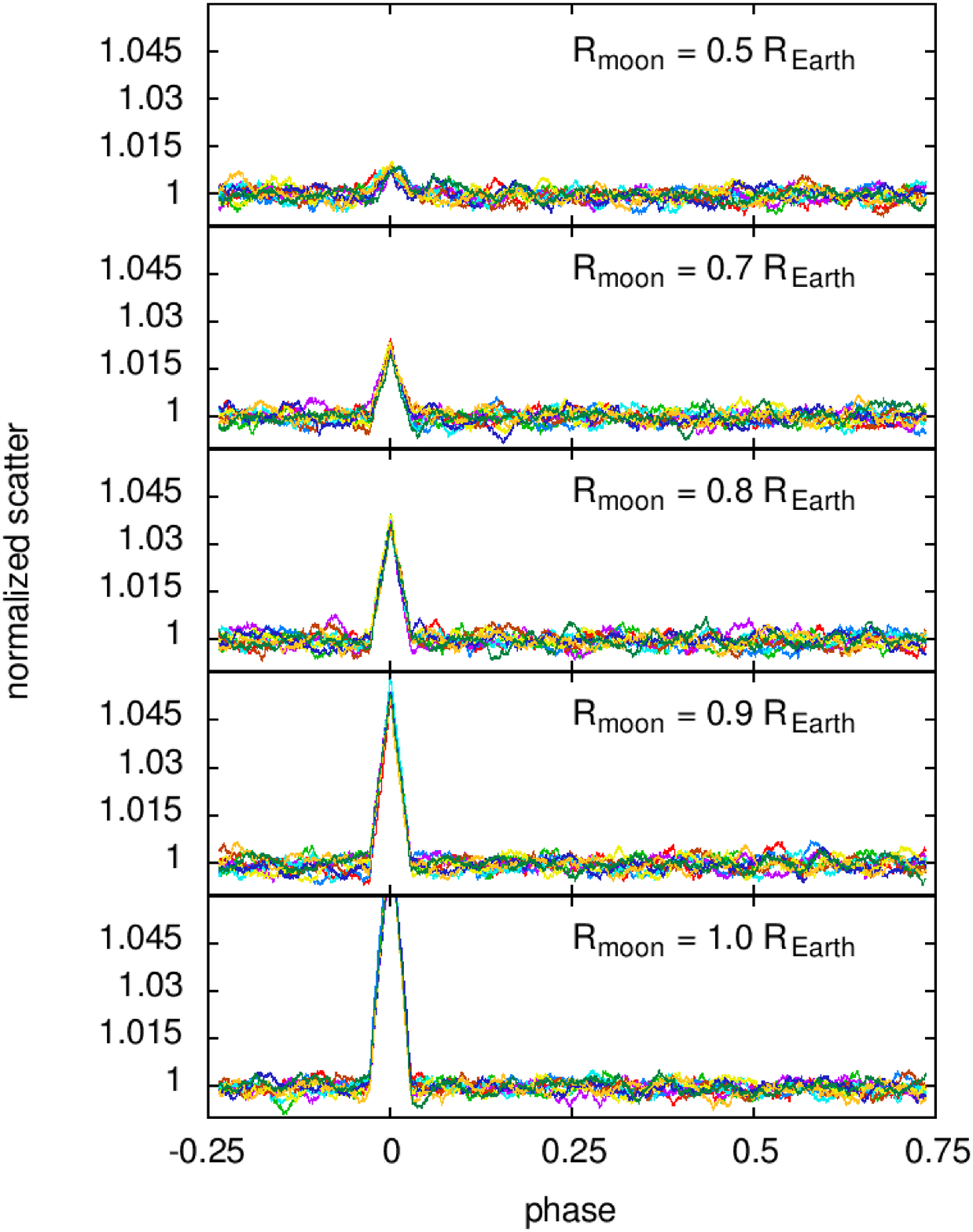,width=5.8cm}
\caption{Normalized scatter peaks due to moon transits in sample
simulations. Each consecutive row shows 10 curves with Kepler short cadence,
long cadence and expected PLATO simulations. Figure lines
show simulations {of an 1 $R_{Jupiter}$ planet with}  a moon of 0.5, 0.7,
0.8, 0.9, 1.0 $R_{Earth}$, respectively.}
\end{centering}
\end{figure*}

\begin{figure*}
\begin{centering}
\epsfig{file=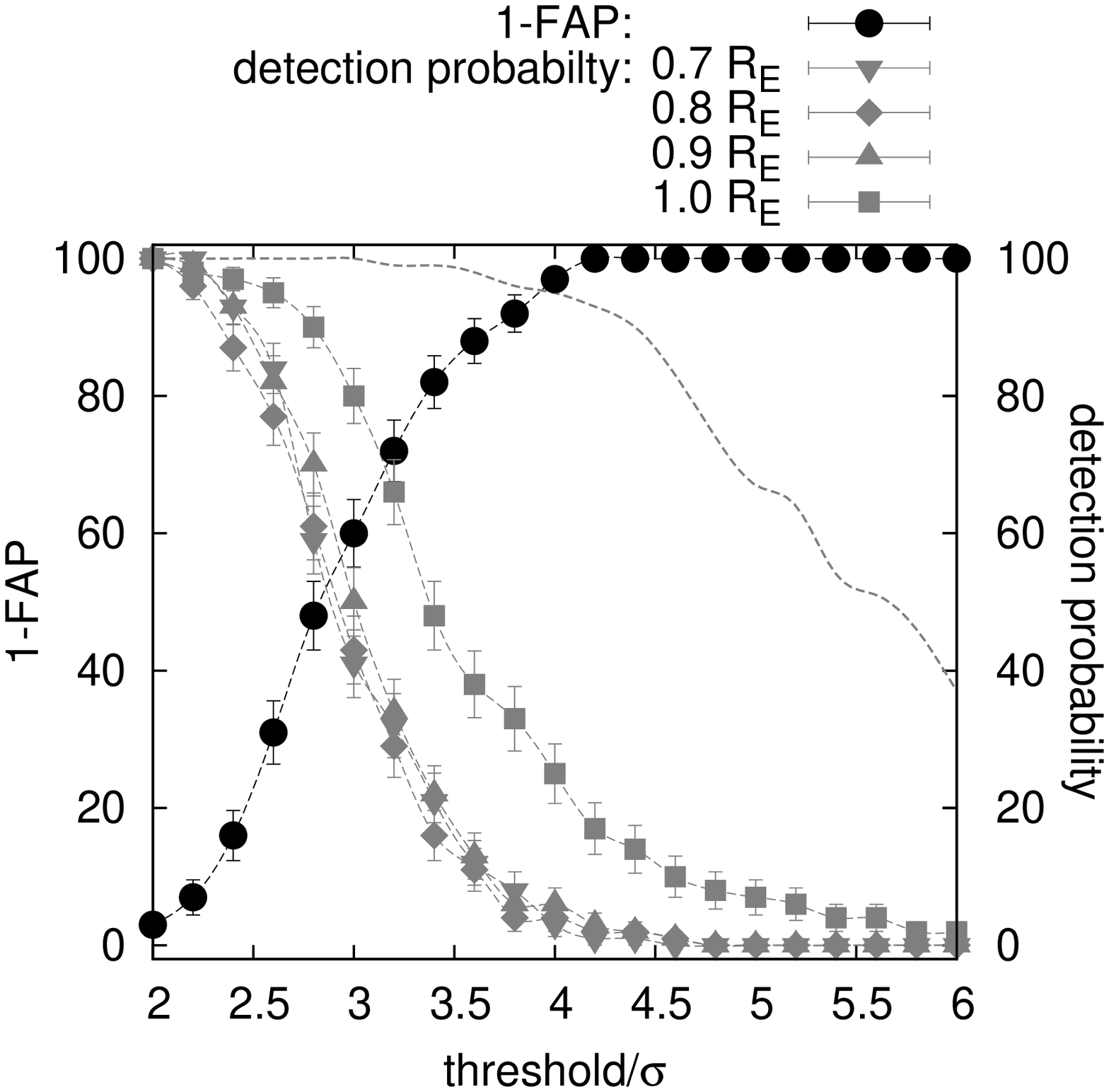,width=7cm,angle=-90}\hskip2cm
\epsfig{file=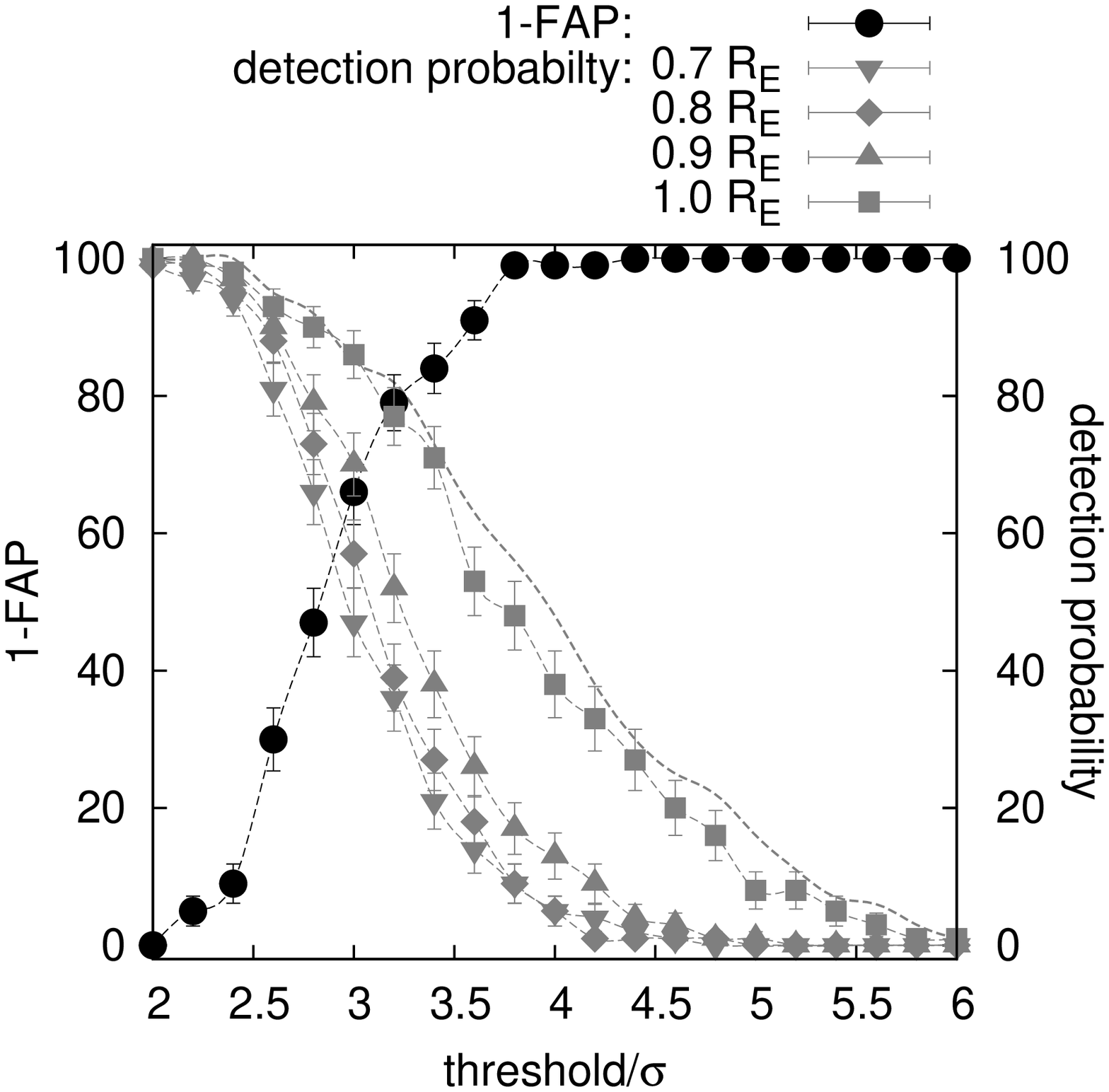,width=7cm,angle=-90}\\
\epsfig{file=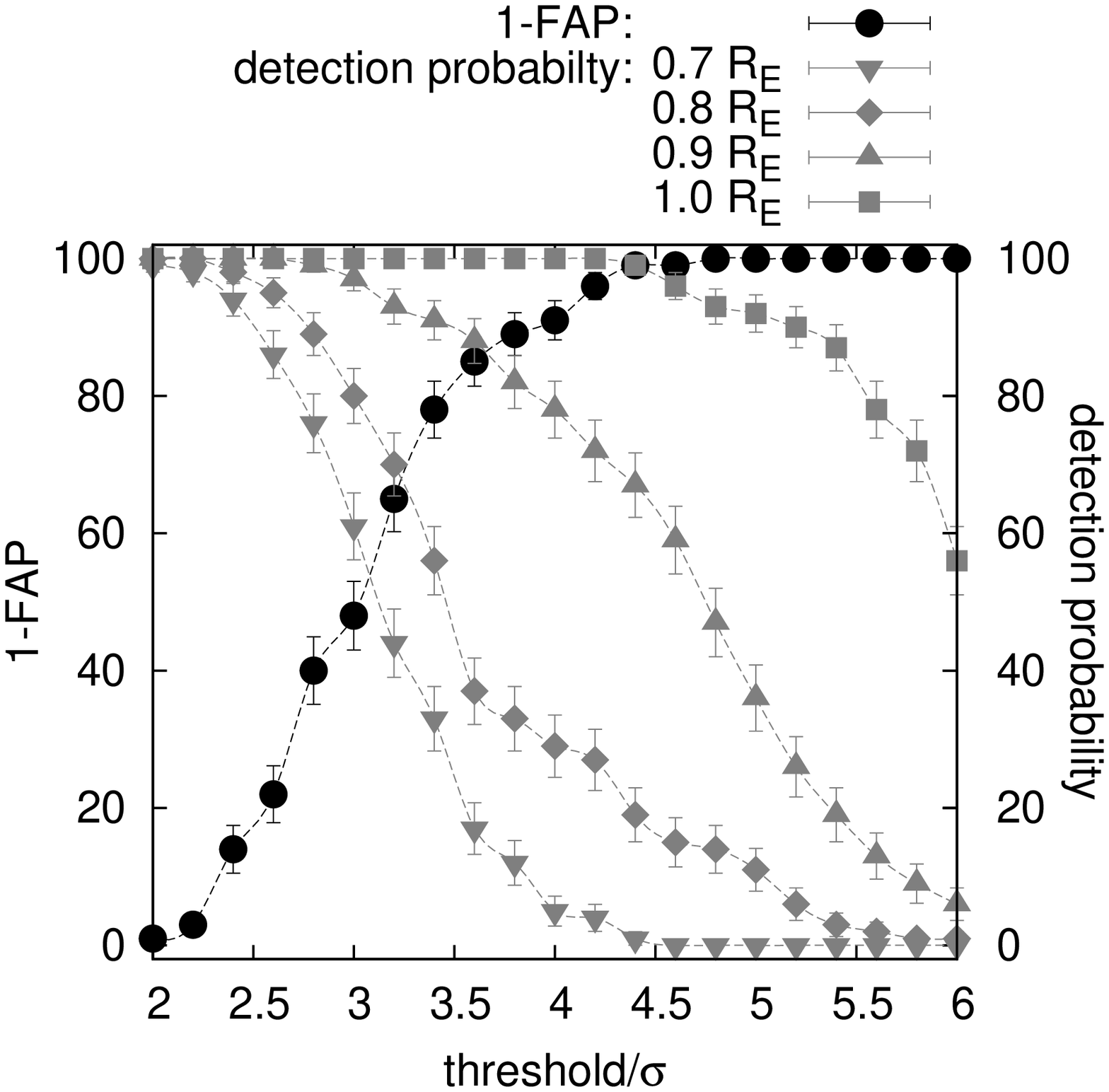,width=7cm,angle=-90}\hskip2cm
\epsfig{file=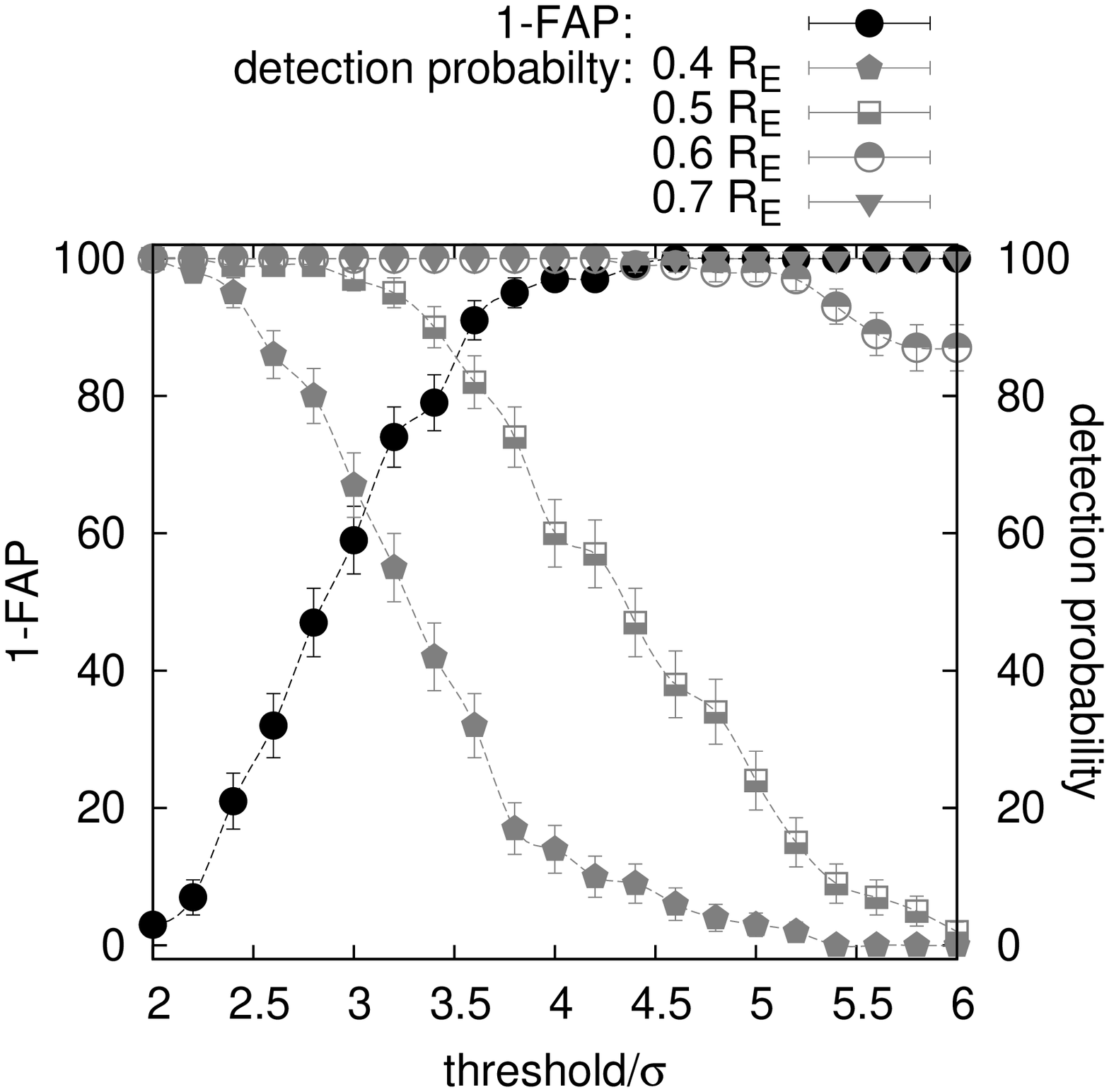,width=7cm,angle=-90}
\caption{Detection probabilities and specificity levels (1 - false alarm
probability) at different threshold levels above the background signal
with $\sigma$ standard deviation. Figures show performance of Kepler long cadence data (top left), best quality ground-based observations (top right), Kepler short cadence data (bottom left) and expected quality of PLATO (bottom right). 
{To illustrate how smearing impairs the detection statistics, in the top panels we plot results from unsmeared reference curves with dashed lines. Smearing does not affect short cadence sampling (bottom panels).}
Different sized moons are colour-coded; note that the change in moon sizes in the bottom right panel.}
\end{centering}
\end{figure*}

\subsection{Recentering}

 {During pre-processing, the transits must be recentered and stretched to have zero TTV and TDV. This is because
there are other sources of TTV and TDV than moons, e.g. perturbing planets. If the folded light curves
are still allowed to exhibit such variation of timings and duration, these events will of course result in 
a Scatter Peak but not due to a moon. However, if TTV and TDV are removed, and the Scatter Peak still survives,
one will have an evidence that there are slight variations in the shape of the transit and in its vicinity. Some other processes can lead to similar results, i.e. the stochastic occultation of individual starspots by a single
planet. In suspicion of some process leading to systemic light curve variations, the variations must be modeled specifically before applying the Scatter Peak evaluation (see also Sect. 4.4). However, it has to be stressed again that most scenarios with a Scatter Peak but without a moon can be excluded by recentering, therefore, this step of pre-processing is the most important ingredient of the method.

In this paper, we used simulations with zero TTV and TDV because the mass of the exomoon was forced to be zero in our light curve simulator. Thus, all detections reflect the photometric effects of the moon itself.}

\subsection{Boxcar size estimation}

The average transit shape is derived from a boxcar median of the folded light curves. 
The length of the boxcar is a sensitive parameter that must be preset with care.
 {Too little boxcars contain too few points, thus the scatter in the folded light curve, and the scatter of the
scatter cannot be determined accurately enough. Too large boxcars, on the other hand,
cover a longer part of the light curve with significant light variation, therefore,
a false Scatter Peak emerges just because the blurred template differs
a lot from the measured light curve. Moreover, the boxcar size will depend on the length
and sampling of data, and on the parameters of the planet.

In every case, the boxcar length must be set manually with numerical experiments. A large number of planet transits must be simulated with the same sampling as the data and varying noise. Then, the largest boxcar must be defined which does not produce a false Scatter Peak with no-moon simulations in the input. 
}

\subsection{Evaluation}

We experienced optimal boxcar lengths of 249, 25, 749, and 1749 photometric points for the 
GB, LC, SC, and PLATO data, respectively. This means that the optimal boxcar was $\approx$400 seconds long,
regardless of the sampling rate. (This boxcar size corresponds to 1/2200 orbital phase.)
Longer boxcars tend to blur the light curve of the planetary
transit too much, while shorter boxcars give too noisy results.
The use of median is necessary because the signal is little, and the possible outliers have to
be eliminated effectively. For such data distributions
(e.g. Gaussian noise with distorted wings), the median is a more 
stable estimate than the mean (Lupton, 1995). We have checked the stability
of our methods utilizing the mean as the local estimate of light curves  and
we indeed experienced that the median is more stable, especially for the length of the
boxcar.

 {In the next step, the median light curve shape must be subtracted from the observations, leading to
the scatter of the light curves (that is partly due to the signal of the moon if it exists). 
The Scatter Peak is in there already, but the data distribution is too noisy for an identification. 
Therefore a smoothing is needed in another boxcar which can be similarly optimized as described above.} 
In our simulations, the second boxcar consisted of 14999, 1499, 44777, and 104999 points
(LC, SC, GB, and PLATO data, respectively); meaning 1.3-times the transit duration.
Surprisingly, so long boxcar is necessary to determine the scatter value with appropriate accuracy.
When the boxcar is centred to the mid-time of the transit, it can measure the
ingress and egress phases of the moon, which may occur well before and well after the 
transit of the planet, depending on the instant geometrical configuration.

If the exoplanet hosts a moon, a well-defined peak of light curve
scatter appears at the phase of the planetary transits.
 {The height of the peak expresses how significantly the scatter will be
increased by systemic light curve distortions. We normalize the height
to the scatter level of the out-of-transit phase.}
The Scatter Peak increases with the size of the moon,
but its height also depends on the quality of data acquired. In Fig 3, a set of
simulations is shown with the different pre-defined data qualities (in successive
columns) and with increasing moon sizes (in successive rows). From this figure 
it can be suspected that as large moons as 1~$R_{Earth}$ can very likely be detected
via the Scatter Peak.

\subsection{Decision making}

In the final step we decide whether there is a convincingly high Scatter Peak in the data.
Even in the no-moon case, the smoothed residuals can mimic a Scatter Peak just by chance,
because of numerical fluctuations. A convincingly high Scatter Peak means such a peak value
which infrequently (False Alarm Probability, $FAP$) evolves from random fluctuations. A Scatter
Peak is convincing if the specificity, $1-FAP$, is close to 1.

The most simple strategy
is to observe whether the observed scatter curve exceeds a pre-set threshold level
at the time of the transit {(i.e., the local scatter is significantly higher than the average scatter plus
a few the scatter of the scatter, which means that the scatter has really increased, and we do not see the result of simple numerical fluctuations).} A lower threshold increases the sensitivity and
decreases the size limit, but the false alarm probability worsens if the value
is set too low. Balancing between sensitivity and specificity sets up the lowest 
appropriate threshold level. To do this, we first decide the specificity of the desired
detection rate, and then simulate and evaluate many (thousands) of no-moon events. The threshold level
belonging to the given specificity is the upper bound of the lowest $1-FAP$ proportion of scatter peaks.
If the observed height of the peak exceeds the threshold level we accept the positive detection.

If one suspects the act of any other process which can lead to a Scatter Peak, this process must be modelled 
and incorporated in selecting a threshold level.  
E.g., in the case of an active star, a spotted stellar model can be fitted.
The null event has to be simulated with this spotted stellar model, 
and the threshold level has to be determined in reference to these light curves.

\subsection{Weighting}

{
A modified implementation of this method involves the appropriate weighting of photometric residuals, instead of
a boxcar smoothing. This will be necessary whenever data of different quality is available. When smoothing in the
boxcar (Sect. 4.4), the expectation for the mean value of the scatter is calculated as, of course,}
\begin{equation}
{\rm sc}\widetilde{\rm at}{\rm ter}_{boxcar} = \sqrt{{1\over N} \sum_{\forall i\ in\ boxcar} r_i^2},
\end{equation}
{where $r_i$ is the residuals inside the boxcar, $N$ is the number of data points, and $\widetilde{\ \ }$ denotes a estimate. 

If the errors of different data points differ, this formulation requires weighting to keep the least-square property of our estimator. In this case the scatter in the boxcar will be estimated as:}
\begin{equation}
{\rm sc}\widetilde{\rm at}{\rm ter}_{boxcar} = \sqrt{\left( \sum_{\forall i\ in\ boxcar} {1 \over \sigma_i^2}\right)^{-1} \sum_{\forall i\ in\ boxcar} {r_i^2 \over \sigma_i^2}},
\end{equation}
{where $\sigma_i$ is the error of the $i$th datapoint. Although Eq. 7 is proportional to the statistic which is usually
tested with $\chi^2$ distribution, we keep suggesting a non-parametric evaluation of the weighted scatter, such as described in Sect. 4.4. This is because $\chi^2$ evaluation assumes that data points come from normal distributions. This is not strictly true in the general case. This improper assumption introduces a little bias, which may easily hide the little signal that we are looking for, or may result in a false alarm. The detection threshold must always be derived from the statistics of the out-of-transit scatter.}

\section{Results}

In Fig 4. we show simulated detection probabilities and specificity ($1-FAP$) estimates, expected for 
ground-based, Kepler long cadence, short cadence, and PLATO-quality simulated observations.
The threshold level is represented in the ordinate, in units of the standard deviation of the
scatter curve belonging to null signal (i.e. out-of-transit). Decreasing curves (in gray colour)
represent the detection probabilities belonging to different size moons, while the black curve
plots specificity. We have deduced that for a clear detection (false alarm
rate $<$1\%{}), threshold levels in the 4.3--4.5$\sigma$ range must be chosen (Fig. 4),
almost independently of the quality of the data (sampling rate and scatter). 

Somewhat surprisingly, top quality ground-based observations promise a 30\%{}
discovery rate for Earth-sized exomoons, having a Scatter Peak above the $4.4\sigma$ 
threshold level. Yet we have not got sub--mmag quality observations of $\approx$100 
full transits of the same planet, but the increasing number of transit observations
and the increasing accuracy of data promises this possibility in the future.

Space-telescopes offer a better detection performance {only with short cadence sampling.}
Selecting $4.4\sigma$ threshold, practically all moons of 1 $R_{Earth}$ size will be
discovered in SC (detection rate is 99\%{}.) The 0.9 and 0.8
Earth-sized moons can be discovered with 70\%{} and 20\%{} in $Kepler$ SC data, respectively.
The detection limit with Kepler is around 0.7 Earth radius. These are such large moons which do not exist
in the Solar System, but they may be found elsewhere. If such moons exist,
they should be discovered in Kepler data, and a possible negative result
will be a significant implication for the lack of so large moons around hot Jupiters.

{Somewhat surprisingly, detection statistics rapidly worsens with longer cadence. We will show that this is primarily a smearing effect (Kipping 2010) rather than a sampling effect. In the top left panel of Fig. 4, we compared the detection statistics with the $Kepler$ LC cadence curves with instantaneous sampling of the unsmeared light curves (1 $R_{Earth}$ size moon; plotted with dashed line), and the smeared light curve (that is the integrated brightness over a the half hour long exposure; plotted with solid squares and error bars). Selecting a 4.4$\sigma$ threshold, the detection rate of our model exomoon would be more than 90\%{} with half hour cadence and without smearing (instantaneous sampling), while it decreases to $\approx$ 15\%{} if smearing is also inculded in the model light curves. The detection statistics of smaller exomoons is identical to the distribution of false positives, so in these cases we do not expect success. The striking impairment of the detections is simply due to the severe smearing on the light curve wings, which blurs the lightcurve of the planet, suppressing the little light variations of the moon itself.}

A real breakthrough is expected by PLATO mission, which is expected to have
significantly lower detection limits (bottom right panel in Fig 4). PLATO should be able
to discover the most exomoons which are larger than 0.6 $R_{Earth}$ with very low
FAP rates. Setting the threshold level to $4.5\sigma$, we expect to discover 40\%{}
of the of the moons with 0.5~$R_{Earth}$ radius, and 7--8\%{} of exomoons with 0.4~$R_{Earth}$.
This experiment will be conclusive in the field of quest for exomoons:
we do know that moons greater than 0.4~$R_{Earth}$ exist: 3 moons in our Solar System
exceeds this size limit.

\subsection{Close-in moons}

\begin{figure}
\includegraphics[angle=270,width=8cm]{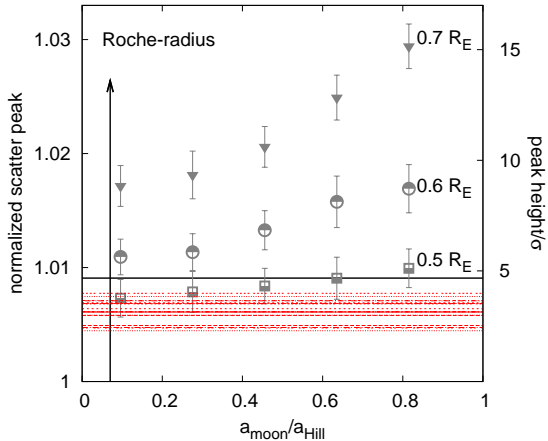}
\caption{Detection statistics of close-in moons. The peaks occurring by random fluctuations are plotted
with various dashed and dotted lines, while the distribution of peak heights in 15 model simulations is plotted by the
symbols with error bars. The left axis plots the normalized height of the peak (level of 1 represents the out-of-transit fluctuations), while the right axis plots the height of the peaks above the out-of-transit fluctuations, scaled by $\sigma$, the standard deviation of the blurred out-of-transit fluctuations. The threshold level of 4.5 times of the out-of-transit scatter is represented by the solid line.}
\end{figure}

 {Besides the size of the moon, the detectability also depends on the orbital radius of the moon.
Light curve effects of close-in moons are limited to the close vicinity of the transit,
shortening the time interval when the light curve distortion is present. This will
decrease the scatter of the residuals, and somehow deteriorate the detection statistics.
(N.B. close-in moons suffer similar observational limitations with the other methods.)
However, the Scatter Peak method is able to detect at least a part of these kind of moons.
To demonstrate this, we illustrate how the detection statistics worsens for a certain
configuration and a single instrument. The complete analysis of close-in moons is a 
complex multiparametric problem and lies beyond the scope of this paper {(see Kipping 2011 for a detailed discussion of a such configuration)}.
}

{We designed systems consisting of the same planet as in the previous simulations, and
systematically decreased the orbital radius of the moons to values of 10\%{}, 28\%{}, 46\%{},
64\%{}, 82\%{} of the Hill-radius. The selected data quality was the PLATO-kind sampling
and noise, while we observed moons of 0.7, 0.6, 0.5 $R_{Earth}$. In Fig. 5, we plot the normalized 
height of the scatter peak, and compare it to the highest peaks by numerical fluctuations
in the no-moon case. The heights of the false alarm peaks are plotted with non-continuous lines, while
the 4.5$\sigma$ threshold is denoted by the solid horizontal line. To the left, we plotted the
Roche-limit, assuming the moon has a densitiy of 3 g/cm$^3$. The symbols show
the detection statistics of the 15 probed configuration. Here, one symbol and the interval lines
represent a whole distribution of detection success. Therefore, the interval covered by the
``error bar'' is the most informative: whenever the error bar goes higher then the threshold level, there
are some correct positive detections above 4.5$\sigma$ level, regardless of the position of the symbol itself.}

 {The plots demonstrate that the detection rate of 0.5 $R_{Earth}$ sized moons decreases with decreasing orbital
distance, however, the ``error bar'' above the solid line expresses that there will be chance for a detection even at 25--30\%{} of the Hill-radius (depending on how lucky configurations occur during the observations). On the other hand, detection success does not decrease for a 0.6 $R_{Earth}$ moon (although the significance does decrease; but all systems are detected,  since all of them remains above the 4.5$\sigma$ threshold). The close-in orbits of moons influence the
detection statistics only for the moons near the size limit of an observational configuration, and does not affect
the detection success of larger moons. The observational bias near the low-end of moon sizes can be determined with
similar numerical experiments, making use of the relevant parameters of the certain system and the quality of the
observations as input.
}

\section{Discussion}

Besides the Scatter Peak, several other methods have been proposed which do have the
potential of discovering an exomoon (e.g. TTV, TDV, Time-of-arrival analysis of pulsars).
A significant limitation for such an application is that the transit configuration
of the planet can also vary because of perturbations. Hence, the detected variations
have to be analysed further and the perturbations must be excluded as the origin of
variations. {Another limitation is the requirement of having $\approx$100 transits
at least for the analysis, which makes the method to be applicable only for planets
with orbital periods of less than 10--20 days. However, this limitation is due to the
planned 3-4-5 year lifetime of space observatories; and we anticipate that it will be
applicable for longer period planets if homogeneous datasets will be available for
some transiting systems. Another possible limitation is of physical nature, i.e. more massive
moons are more rapidly removed by tidal forces (see Barnes and O'Brien 2002 for a
detailed description), while moons beyond
$\approx$ 0.5 of the Hill-radius are not stable on the long time scale (several billion years in some cases,
Domingos et al. 2006).}

In comparison to these methods, Scatter Peak analysis will be insensitive to perturbation
effect in a slightly modified form. If each individual transit is re-centered to the
estimated position of the planet, the local fluctuations caused by
the timing variations will be eliminated.  
 {The exact implementation would be ``planetocentric'' recentering, 
for that barycentric TTV is only a good proxy if there is a moon. But in fact, any recentering
method may be applied here: if there is a moon and we get a positive detection we got what
we were looking for. If there is no moon, parametric transit time and the light curve photocenter
(Szab\'o et al. 2006) coincides exactly, and any of them is appropriate to eliminate planet
perturbation effects.} The varying position of the moon, however,
will still result in systematic variations of the light curve, which will increase the
local scatter and hence, enable to infer that a moon is present. Another advantage is 
the non-parametric nature of the Scatter Peak method, which warrants that {\it a priori}
assumptions of the shape of the transit light curve do not influence the result. 
We remark that 
the method is now tested for different moon sizes and orbital radii, while a more general
testing (also for inclined and non-circular orbits) is the task of a
forthcomming paper. However, the current level of testing does not influence
the suggestion that the Scatter Peak method can help a lot in discovering
the exomoons.

ESA's planned PLATO mission will offer a great opportunity for the detection of exomoons,
because of the large number of the targeted stars ($\approx 250,000$) and their favourable
brightness (8--11 mag). These stars will be mostly cool dwarfs, hence there will be a good
possibility for very accurate radial velocity measurements, and to observe the Rossiter-McLaughlin
effect in transit. It may be possible to detect the moon in the Rossiter-Mclaughlin effect,
too, as a confirmation of the moon, which is independent from the Scatter Peak (Simon et al. 2010).
Kaltenegger (2010) suggested that even the atmosphere of an Earth-sized exomoon can be
detected, which is the most important if such an exomoon orbits in the habitable zone
(Kipping et al., 2009).

The most significant error source for the Scatter Peak method is the quality of de-trending the
observed light curves. 
For space photometry, the slowly changing zero-point of the data could be a significant limiting factor, because removing the instrumental trends is everything but trivial.
Also, if the de-trending of Earth-based photometry involves a comparison of the observation to a 
set of parametric templates, there will be risk that the algorithm will try to interpret 
the signal of the moon as systematics and eliminate its signal.

In summary, we conclude that testing the Scatter Peak from a sequence of light curves is
a promising tool for detecting moons directly in the light
curves. The success relies on three important
conditions:

\begin{itemize}

\item{} All light curves must be stacked in such way that
the transit time of the planet exactly coincide
in each of the analysed light curves.
\item{} Transit observations must include the out-of
transit phases before and after the transit of the
planet, where the scatter due to the moon is the
highest. The wings must span at least as long as
the transit duration.
\item{} Trend filtering of the light curves must be carried out 
in such a way that small deviations immediately before and after 
the transit of the planet shall remain unaffected.

\end{itemize}

\section*{Acknowledgments}

This project was supported by 
the Hungarian OTKA Grants K76816 and MB08C 81013, and the ``Lend\"ulet'' Young
Researchers' Program of the Hungarian Academy of Sciences.
We thank the referee D. M. Kipping for valuable comments, that helped to impove the paper.

\section{Deriving the projected position of the moon}

We know that $\nu$ follows a uniform distribution and the question is the density function of $x=\sin\nu$ (we assume length is measured in the unit of $a_m$ for the sake of simplicity). Let $F(\nu)=p(\nu '>\nu)$ and $F(x)=p(x' > x)$ be the cumulative distribution of the same set of positions, parametrized by $\nu$ and $x$, respectively (here $p$ represents probabilities, $\nu'$ and $x'$ are generic running parameter). Since $\nu$ is uniformly distributed,
\begin{equation}
{dF(\nu) \over d\nu} = {1\over \pi}
\end{equation}
where $\nu$ is element of the interval $[-\pi/2,\pi/2]$. By the assumption made on the length scale, $x_m=\sin\nu$. The differentials of this transformation are $dx_m = \cos \nu d\nu$, and because $\cos\nu = \sqrt{1-x_m^2}$, 
$d\nu = dx_m / \sqrt {1-x_m^2}$. Substituting $d\nu$ with $dx$ leads to the result,
\begin{equation}
{dF(\nu) \over dx} = {1\over \pi \sqrt{1-x^2}}
\end{equation}
which is Eq. 1 of the paper after $a_m$ is written explicitly to represent the appropriate length scale.
Although the expression is singular at the end points ($x/a_m=\pm1$), there is no problem with its interpretation because the expression can be integrated everywhere (the question "What is the probability of having the moon in the outermost 1\%{} of the projected orbit?" can be answered exactly, despite the singularity.)




\end{appendix}

\label{lastpage}

\end{document}